\documentclass[11pt]{article}
\usepackage{setspace}
\usepackage{amssymb}
\usepackage{amsmath}
\usepackage{amsfonts}
\usepackage{bm}

\def\be{\begin{equation}}

\def\ee{\end{equation}}

\newcommand\vecbf[1]{{\bf #1}}
\newcommand\mat[1]{ {\mathbb #1}}

\def\dd{\partial}

\newcommand\of[1]{\left( #1 \right)}
\newcommand\sqof[1]{\left[ #1 \right]}

\def\bea{\begin{eqnarray}}

\def\eea{\end{eqnarray}}

\newcommand\eps{\epsilon}
\newcommand\refeq[1]{(\ref{#1})}
\begin{document}

\singlespace

\begin{flushright} BRX TH-657 \\
CALT 68-2876
\end{flushright}

\vspace*{.3in}

\begin{center}

{\Large\bf  Symmetrically reduced Galileon equations and solutions}

{\large S.\ Deser}

{\it Physics Department,  Brandeis University, Waltham, MA 02454 and \\
Lauritsen Laboratory, California Institute of Technology, Pasadena, CA 91125 \\
{\tt deser@brandeis.edu}
}

{\large J.\ Franklin}

{\it  Reed College, Portland, OR 97202 \\
{\tt jfrankli@reed.edu}}

\end{center}

\begin{abstract}
The maximally complicated arbitrary-dimensional ``maximal" Galileon field equations simplify dramatically for symmetric configurations.  Thus,
spherical symmetry reduces the equations from the $D-$ to the two-dimensional Monge-Ampere equation, axial symmetry to its cubic extension etc. 
We can then obtain explicit solutions, such as spherical or axial waves, and relate them to the (known) general, but highly implicit, lower-$D$ solutions.
\end{abstract}

\section{Introduction}
Galileons, while popular in the cosmological literature, deserve to be better understood on their own. The present contribution is the study  
of their ``maximal" forms in arbitrary dimension $D$, with a view to finding some interesting simplifications and explicit solutions of their, rather exotic, field equations. Such searches are usually most successful for systems with symmetry, as turns out to be especially true here. We will find dramatic reduction both of the effective equations and their solutions as the degree of symmetry is increased: Thus, spherically symmetric configurations obey the $D=2$ Monge-Ampere equation rather than the $D$-dimensional Hessian one. This enables us to display quite general classes of explicit solutions, independent of the models' dimensionality, as against the known general but quite implicit  ones.

We begin with a summary of the models; in the formulation of~\cite{DDE}, the actions for the scalar field $\phi$ are given by
\begin{equation}
I = \int d^D x \, \phi \, \left[ \eps^{\mu\nu\ldots \alpha} \, \eps^{\gamma\delta \ldots \beta} \, \phi_{,\mu\gamma} \, \phi_{,\nu\delta} \ldots \, \phi_{,\alpha \beta}\right] = \int d^D x \phi \, \det(\phi_{,\alpha \beta} ).
\end{equation}
The field equation states that the $D$-dimensional determinant of its second derivatives -- i.e., the Hessian~\cite{Hesse} vanishes:
\begin{equation}
\det(\phi_{,\alpha \beta}) = 0.
\end{equation}
As already noted in~\cite{Hesse}, all $\phi$ that are independent of at least one Cartesian space-time coordinate are trivially solutions: vanishing of any derivative annihilates a complete row and column of the determinant. This implies a sort of anti-Birkhoff theorem: since all static functions are solutions, interesting ones must depend on $t$. It also demonstrates the fact that one can have too much symmetry, here that of a cyclic coordinate. [Note also that unlike for normal Lorentz invariant systems, arbitrary constant rescaling of each coordinate separately is allowed because each term in the determinant expansion contains all derivatives.] We now turn to the reduction of (2) effected by less drastic symmetries.

\section{Spherical Symmetry}
We begin with maximal, spherical, symmetry, $\phi=\phi(r,t)$ and show that $\phi$ obeys the ($D=2$) Monge-Ampere equation for all $D$. [Missing angular dependence is not the same as missing Cartesian coordinates--in the former case, all the $x^i$ are still present, just in a particularly symmetric combination.]   Note first that
\begin{equation}
\phi_{,i}= x_i/r\,  \phi' \, \, \, \, \, \, \, \, \, \phi_{,ij}=\delta_{ij}  \, (\phi'/r) + x_i \, x_j /r^3 (\phi ''\, r - \phi')
\end{equation}
where prime$=d/dr$, dot$=d/dt$.   We write $\phi_{,\mu\nu}$ as the matrix
\begin{equation}\label{Hessentry}
\phi_{,\mu\nu} \dot = \left( \begin{array}{cc} \ddot \phi & \dot \phi' \, x_i/r \\
\dot \phi' \, x_i/r & \delta_{ij} \, \phi'/r + x_i \, x_j/r^3 \, [\phi'' \, r - \phi']
\end{array}\right).
\end{equation}
This implies that the full Hessian matrix~\refeq{Hessentry} is of the form
\begin{equation}
\mat S  \equiv \left( \begin{array}{cc} \ddot\phi & \vecbf b^T \\ \vecbf b & \mat D \end{array} \right),
\end{equation}
with determinant
\begin{equation}\label{blockdecomp}
\det(\mat S) = \ddot\phi \, \det \left[ \mat D - \vecbf b \, \ddot\phi^{-1} \, \vecbf b^T \right],
\end{equation}
the standard result for block matrices.  Using $\vecbf b_i = \dot \phi' \, \frac{x_i}{r}$ and $\mat D_{ij} = \frac{\phi'}{r} \, \delta_{ij} + x_i \, x_j \, \frac{\phi'' \, r - \phi'}{r^3}$, the matrix in~\refeq{blockdecomp} is \begin{equation}
\sqof{\mat D - \vecbf b \, \ddot\phi^{-1} \, \vecbf b^T }_{ij} = \delta_{ij} \, \frac{\phi'}{r} + x_i \, x_j \, \left[ \frac{\phi'' \, r - \phi'}{r^3}  - \frac{\dot \phi'^2}{r^2 \, \ddot \phi} \right].
\end{equation}
But Sylvester's theorem states that
\begin{equation}
\det\left[ \alpha \, \delta_{ij} + \beta \, x_i \, x_j \right] = \alpha^{D-1} \, \of{1 + \frac{\beta}{\alpha} \, r^2},
\end{equation}
hence
\begin{equation}
\begin{aligned}
\det \left[ \mat D - \vecbf b \, \ddot\phi^{-1} \, \vecbf b^T \right] &= \sqof{\frac{\phi'}{r}}^{D-1} \, \of{1 + \frac{r}{\phi'} \, \of{ \phi'' - \frac{\phi'}{r} - \frac{\dot \phi'^2}{\ddot \phi}}} \\
&= \sqof{\frac{\phi'}{r}}^{D-2} \, \of{\phi'' - \frac{\dot \phi'^2}{\ddot \phi}}.
\end{aligned}
\end{equation}
By~\refeq{blockdecomp}, and dropping the overall coefficient, the reduced field equation is 
\begin{equation}\label{HPDE}
  \of{ \phi'' \, \ddot \phi - \dot \phi'^2} = 0,
\end{equation}
independent of $D$, as promised. In the most degenerate, one-variable case $\phi=\phi[\sqrt{x^2}=\sqrt{t^2\pm r^2}]$, the equation reduces to $\phi''=0$, whose solution is just $\phi =a\, \sqrt{x^2} + b$; this is of course distinct from the ``pure gauge" $\phi\sim a_\mu\, x^\mu+b$, whose second derivatives vanish identically.  This solution exemplifies the result~\cite{Fairlie} that any function homogeneous of order one in $(t,r)$ is a solution.

It is not difficult to guess some classes of interesting solutions of
(10). In particular, there are traveling solutions, $\phi(t,r)= \Phi(a\, t+b\, r)$.
Indeed, superpositions of traveling solutions, $\phi = \Phi_1(a \, t + b \, r) + \Phi_2(c \, t + d \, r)$ seem to exist, subject ``only"
to the condition $a\, d=b \, c$; unfortunately, this condition reveals the superposition to be the initial $\phi=\Phi(a \, t + b \, r)$ in disguise:  recalling
the freedom of rescaling $(t,r)$ by arbitrary constants, we choose
to remove $a$ and $b$ from $\Phi_1$, which then makes $\Phi_2 = \Phi_2 (c/a\, t + d/b \, r)$;
 but since $a \, d =b \, c$,  $\Phi_2$ is again a function of just $t+r$, like the
(rescaled) $\Phi_1$ and $\phi=\Phi(t+r)$.  Our traveling wave broadens, in a sense, the homogeneity theorem of~\cite{Fairlie}: obviously, any -- even independent -- rescalings of $t$ and $r$ are permitted for $\Phi(a\, t+b\, r)$, as indeed for their lower symmetry extensions studied below.

Actually, the general, if highly implicit, solution of~\refeq{HPDE} is known and can be correlated to some of the explicit ones; we follow~\cite{Chaundy, Fairlie}.  The idea is to generate $\phi(t,r)$ whose $\dot \phi$ and $\phi'$ are related -- then the columns of the Hessian matrix will not be independent, and hence the determinant of the matrix will vanish.  One can accomplish this by setting $\dot \phi(t,r) = f(u)$, $\phi'(t,r) = g(u)$, where $f$ and $g$ are some functions of the single variable $u(t,r)$.  For then
\begin{equation}
\ddot \phi = f'(u) \, \dot u \, \, \, \, \, \, \, \, \, \phi'' = g'(u) \, u' \, \, \, \, \, \, \, \, \, \dot \phi' = f'(u) \, u' = g'(u) \, \dot u,
\end{equation}
and the combination in~\refeq{HPDE} vanishes.  In order to generate a $\phi$ with the appropriate derivative relations, we take
\begin{equation}\label{phiis}
\phi(t,r) = t \, f(u) + r \, g(u) + u \, a,
\end{equation}
for an arbitrary constant $a$.  Then
\begin{equation}
\dot \phi = f + \dot u \, (t \, f' + r \, g' + a) \, \, \, \, \, \, \, \, \, \phi' = g + u' \, (t \, f' + r \, g' + a) 
\end{equation}
so that $\dot \phi = f$ and $\phi' = g$ hold if
\begin{equation}\label{auxis}
t \, f'(u) + r \, g'(u) + a = 0.
\end{equation}
The equations~\refeq{phiis} and~\refeq{auxis} must both be satisfied, and together, implicitly define the most general solution to~\refeq{HPDE}.  The freedom in the choice of $f$ and $g$ can be used to reproduce explicit solutions.  Perhaps the easiest of these is the single-variable one found earlier:  taking $(f,g) = (\cos{u}, \sin{u}) = (t,r)/\sqrt{r^2 + t^2}$ yields the desired $\phi = \sqrt{t^2 + r^2}$; hyperbolic functions yield $\phi = \sqrt{t^2 - r^2}$.

\section{Lower symmetries}

 It is rather clear what to expect as the symmetry level decreases. Take first axial symmetry, with dependence on one axial, $z$-direction, and
the radial variable $\rho^2= x_1^2 + \ldots + x_ {D-2}^2$, so $\phi=\phi(t, \rho,z) \equiv \phi(Y_i)$. We omit the details, but from the spherical discussion, one expects the 
$D=3$ determinant extension of Monge-Ampere,
\begin{equation}
 \hbox{det}\of{\frac{\dd^2 \phi}{\dd Y^i \, \dd Y^j}} = 0.
\end{equation}
This is indeed borne out by Sylvester-like decomposition of the full $\phi_{,\mu\nu}$ into a $D=2$ $(t,z)$ times the ``planar" spherically symmetric $D-2$.

The solutions are again obtainable in a general, but implicit form.  It is perhaps not surprising that $\phi = a \, \sqrt{t^2 \pm \rho^2 \pm z^2} + b$, and $\phi = \Phi(a \, t + b \, \rho + c \, z)$ are also solutions.  In surprising contrast to the spherical result, however, we find here that superposition, $\phi = \Phi_1(a \, t + b \, \rho + c  \, z) + \Phi_2(d \, t + e \, \rho + f \, z)$ holds: unlike the spherical case, there is no constraint among the constants.

Finally, we derive the effective reduced Hessian equations for arbitrary symmetry configurations, by writing (2) in terms of the appropriate generalized coordinates. Let us call, collectively, the respective relevant and missing directions ``$\vecbf r$" and ``${\bm \theta}$", boldfaces indicating that there can be more than one. The number $R$ of 
$\vecbf r$-directions determines the dimension of the reduced, $(t,\vecbf r)$, Hessian. Generalized coordinates of course require introduction of covariant derivatives, replacing $\dd_i \, \dd_j$ by $D_i\,  \dd_j$ on our scalar. [The $D_i$ commute in flat space, and there is no change in the time derivatives.]  The Hessian equation then becomes
\begin{equation}
\eps^{t \vecbf r {\bm \theta}}\,  \eps^{t \vecbf r {\bm \theta} } D\, \dd \, \phi   \ldots D \, \dd \, \phi =0. 
\end{equation}
Its dimension is still $D$, but effectively contracts after we implement the symmetries: The factors $D_i\, \dot \phi$ reduce to $\dd_{\vecbf r} \, \dot\phi$. The  $D_{\vecbf r} \dd_{\bm \theta}$ or $D_{\bm \theta} \,  \dd_{\vecbf r}  \sim \Gamma^{\bm \theta}_{\vecbf r {\bm \theta}} \,  \dd_{\bm \theta} \phi$ vanish: the only non-zero connections have two ${\bm \theta}$ and one $r$, while $\dd_{\bm \theta} \phi=0$; also, $D_{\vecbf r} \dd_{\vecbf r} = \dd_{\vecbf r} \, \dd_{\vecbf r}$. There remains 
$D_{\bm \theta} \,  \dd_{\bm \theta}  \, \phi \sim \Gamma^{\vecbf r} _{\bm \theta  \theta} \,  \dd_{\vecbf r} \phi \sim \dd_{\vecbf r} \phi$ multiplied by $r$ and an angular coefficient. Further, $D_{\bm \theta} \, \dd_{\bm \theta} \, \phi$ is a common factor to every term of (16): it provides the (only) non-vanishing contraction of the ${\bm \theta}$ indices in $\eps \eps$; precisely because it is common to each term in (16), it may be canceled out of it. So (16) reduces, as predicted, to a determinant of dimension $R+1$, (there are of course  $D-2-R$ ${\bm \theta}$), formally like our spherical result (10), but now the vector $\vecbf r$ determines the number of $\phi$ factors there: quadratic for spherical, cubic for axial, symmetries, etc. Specifically, in our vector notation, the final equation is just 
\begin{equation}
 \left[ \ddot \phi  \, \dd_{\vecbf r} \, \dd_{\vecbf r} \, \phi   -  \dd_{\vecbf r} \dot \phi  \dd_{\vecbf r} \, \dot \phi\right] =0,     
\end{equation}
whose $(R+1)$ dimension is implicit in our vector notation.
\section{Summary}
We have been able to tame the general $D$, maximal Galileons for configurations with arbitrary degrees of symmetry, thereby gaining some concrete view of their explicit behavior beyond the, rather forbidding, implicit general solutions of the full Hessian equation. Similar reductions could be envisaged in terms of, e.g., light-cone coordinate choices where $t$ becomes mixed with spatial variables. Ours can hardly pretend to be a detailed analysis, however. We have not attempted to seek solutions with proper asymptotic behavior, let alone study the necessarily non minimally coupled Galileon-gravity equations -- a qualitatively harder challenge, if perhaps still feasible in the spherical limit.

\section{Acknowledgments}

SD thanks D Fairlie for communicating unpublished results and pointing out [3], and R Palais for providing a glimpse of the enormous Hessian literature. SD was supported in part by NSF PHY-1064302 and DOE DE-FG02-164 92ER40701 grants.


\begin{thebibliography}{99}
\bibitem{DDE} C. Deffayet, S. Deser, and G. Esposito-Farese, {\it Phys. Rev. D} {\bf 80} 064015 (2009), arxiv:0906.1967.
\bibitem{Hesse} O. Hesse, {\it J. Reine Angew. Math.} {\bf 41} 285 (1851). 
\bibitem{Chaundy} T. W. Chaundy, ``The Differential Calculus", Oxford University Press, 1935.
\bibitem{Fairlie} D. B. Fairlie and A. N. Leznov, {\it J. Geom. Phys.} {\bf 16} 385 (1995), hep-th/9403134;
D. B. Fairlie, {\it J. Phys. A}, {\bf 44} 305201 (2011), arxiv:1102.1594.
\end{thebibliography}
\end{document}